\documentstyle[11pt,epsfig]{article}

 \renewcommand{\theequation}{\arabic{section}.\arabic{equation}}

\newcommand{\beq}{\begin{equation}}
\newcommand{\eeq}{\end{equation}}
\newcommand{\bea}{\begin{eqnarray}}
\newcommand{\eea}{\end{eqnarray}}

\def \pa {\partial}

\def \ha{\widehat}

\def \ra {\rightarrow}

\def \La {\Lambda}
\def \Da {\Delta}
\def \b {\beta}
\def \a {\alpha}

\def \Ga {\Gamma}
\def \ga {\gamma}

\def \da {\delta}

\def \om {\omega}

\begin{document}
\begin{titlepage}

\begin{flushright}
BA-TH/01-415\\
CERN-TH/2001-149 \\
hep-th/0106019
\end{flushright}

\vspace*{3mm}

\begin{center}
\huge
{\bf  Localization of Scalar Fluctuations \\
in  a Dilatonic Brane-World Scenario}
\vspace*{1cm}

\large{
V. Bozza\footnote{valboz@sa.infn.it}${}^{(a,b)}$,
M. Gasperini\footnote{gasperini@ba.infn.it}${}^{(c,d)}$ and
G. Veneziano\footnote{venezia@nxth04.cern.ch}${}^{(e,f)}$}

\bigskip
\normalsize
${}^{(a)}$
{\sl Dipartimento di Fisica ``E. R. Caianiello", Universit\`a di Salerno, \\
Via S. Allende, 84081 Baronissi (SA), Italy\\
\vspace{0.2cm}
${}^{(b)}$
INFN, Sezione di Napoli,
Gruppo Collegato di Salerno, Salerno, Italy} \\
\vspace{0.2cm}
${}^{(c)}$
{\sl Dipartimento di Fisica, Universit\`a di Bari, \\
Via G. Amendola 173, 70126 Bari, Italy\\
\vspace{0.2cm}
${}^{(d)}$
INFN, Sezione di Bari,
Bari, Italy} \\
\vspace{0.2cm}
${}^{(e)}$
{\sl Theoretical Physics Division, CERN, CH-1211 Geneva 23, Switzerland \\
\vspace{0.2cm}
${}^{(f)}$
Laboratoire de Physique Th\'eorique, Universit\'e Paris
Sud, 91405 Orsay, France}


\vspace*{5mm}

\begin{abstract}
We derive and solve the full set of scalar perturbation equations
for a  class of $Z_2$-symmetric five-dimensional geometries generated  
by a bulk cosmological constant and by a 3-brane non-minimally coupled  
to a bulk dilaton field. The massless scalar modes,  like their tensor
analogues, are localized on the brane, and provide long-range
four-dimensional dilatonic interactions, which are generically present
even when matter on the brane carries no dilatonic charge. The
shorter-range corrections induced by the continuum of massive scalar
modes are always present: they persist even in the case of a
trivial dilaton background (the standard Randall--Sundrum
configuration) and vanishing dilatonic charges.
\end{abstract}

\end{center}
\end{titlepage}


\section{Introduction}
\setcounter{equation}{0}
\def\theequation{\thesection.\arabic{equation}}

The possibility, first discovered in the context of Horava--Witten (HW)  
heterotic M-theory \cite{1,2}, that  our Universe  could lie
on a hypersurface (a ``brane") embedded in some higher-dimensional
``bulk" space-time --the so-called brane-world scenario-- has recently  
attracted considerable attention. In the original HW paper, our world is  
a 9-brane sitting at one of the boundaries of an eleven-dimensional bulk, 
while  in a large class of
M-theory models \cite{4,5}, in which  six dimensions are compactified in  
a more traditional Kaluza--Klein (KK) way, one can envisage  
constructing fully consistent four-dimensional brane-world scenarios  
with, effectively, a five-dimensional bulk.
Unfortunately, finding brane configurations which are
consistent with all stringy constraints has proved to be a very
hard, if not impossible, task.

At a more phenomenological level, i.e. when  problems with
quantization of the higher-dimensional gravity theory are ignored, one  
 can consider the dimensions orthogonal to the 3-brane as either
compact   and large \cite{9,6,7}, or as having infinite proper size
\cite{3,8,9a}. In  
 the latter (so-called Randall--Sundrum (RS)) case,  the bulk geometry  
 is bent
by an appropriate ``warp-factor",  providing a crucial difference
with respect to the ``old" KK scenario, where the bulk
geometry is simply the direct product of an ``internal" and an
``external" manifold. In particular, unlike the KK models, RS models  
are able to  reproduce the four-dimensional Newton law at large
distances on the brane by {\it dynamically} binding the massless
gravitons to it \cite{9a}.

As a consequence of the non-factorized structure of the metric in
RS-type scenarios,  previous
approaches to the study of metric fluctuations  in higher-dimensional
backgrounds \cite{10}--\cite{15}, based on the isometries of a
factorizable geometry, cannot  be applied directly to the brane-world
scenario. A new gauge-invariant formalism is required, like the one
developed in \cite{16}. The classical
and quantum analysis of metric fluctuations is of
primary importance for understanding the possible localization of
massless modes on the brane, as well as the nature of the  orrections
to   the long-range interaction due to the continuum of massive modes
that   are typically living in the bulk. Until now, the study of this
problem   has been mainly focused on the structure of tensor (i.e.
transverse-traceless) perturbations of the
 bulk geometry (see \cite{17} for a general discussion).

In all string/M-theory models, however, the graviton enjoys the 
company of   (perturbatively) massless scalar partners (the dilaton,
compactification   moduli, etc.). These typically induce long-range 
interactions of gravitational strength \cite{TV}, and are therefore
dangerous in view   of the
existing experimental tests (see for instance \cite{tests}).  The
standard   way to solve this problem is to assume that these scalars get
a SUSY-breaking non-perturbative mass (for alternatives see e.g.
\cite{DP}). However, one may ask whether RS-type scenarios can offer an
alternative solution to (or an alleviation of) this problem, e.g. by
{\it not} confining scalar fluctuations on the brane. If this were the  
case, scalar interactions on the brane would be suppressed, or possibly  
become short-ranged, and brane-world scenarios would naturally solve one  
of the most serious potential problems with higher-dimensional or
stringy extensions of the Standard Model and  General Relativity. This  
particular aspect of brane-world scenarios has never been completely
addressed, to the best of our knowledge,  in
spite of many studies recently appeared in the literature, and devoted
to the perturbations of a brane-world background \cite{18}--\cite{26}.

In this paper we present a detailed discussion of the localization of
scalar metric fluctuations in a typical example of brane-world
scenario. Unlike gravitons, which are decoupled from matter
fluctuations, the scalar fluctuations of the bulk geometry  are in
general non-trivially coupled to the matter sources.
We shall thus consider a
non-compact, $Z_2$-symmetric, five-dimensional background,
generated by a positive tension 3-brane and by a bulk dilaton field
coupled to the brane and to the (negative) bulk energy density.  We
shall restrict ourselves, in particular, to the gravi-dilaton  
solutions discussed in
\cite{27} (hereafter called CLP backgrounds, for short), which generalize
the AdS$_5$ RS scenario \cite{9a} in the presence of a bulk
scalar field, and which are already known to guarantee the localization
of tensor metric fluctuations \cite{27}.

By extending the analysis of the perturbations, and by using an
appropriate gauge-invariant formalism \cite{16}, we find that the same
class of CLP backgrounds that localize massless tensor perturbations on the
four-dimensional brane also  localize massless scalar perturbations
(i.e. the dilatonic interaction). However, the short-range corrections to
the scalar interactions, due to the massive modes propagating in the
extra-dimension, are in general different from the higher-dimensional
corrections affecting the pure tensor part of the gravitational
interaction.
We also find that, in general,  scalar metric fluctuations
exhibit a non-trivial, ``self-sustained" spectrum of solutions,
even for a trivial dilaton background. This
implies that long- and short-range gravitational interactions on the  
brane are
effectively of the scalar-tensor type, in agreement with previous
results \cite{27a}, and are therefore subject to strong phenomenological
constraints.

The paper is organized as follows. In Section 2 we present the action
and the classical equations of motion for a 3-brane
non-minimally coupled to a five-dimensional gravi-dilaton background,
and we retrieve the whole class of CLP solutions \cite{27}. In Section
3 we give the full set of scalar perturbation equations in the so-called
``generalized longitudinal gauge" \cite{16}, and we find the four
independent canonical variables diagonalizing  them.
 In Section 4 we discuss the localization of
massless modes, and determine the class of backgrounds admitting
long-range dilatonic interactions confined on the brane. In Section 5 we
present the general spectrum of solutions for the massive modes
that propagate throughout the bulk, and  determine the
relative magnitude of  their amplitudes. In Section 6 we evaluate, in the
weak field limit,  the leading-order corrections to the effective
gravitational potential generated by a static source with a point-like
mass and dilatonic charge, confined on the brane.
The main results of this paper are finally summarized in
Section 7.

\section{Background equations}
\setcounter{equation}{0}
\def\theequation{\thesection.\arabic{equation}}

We shall consider a five-dimensional scalar-tensor
background $\{g_{AB}, \phi\}$, possibly arising from the bosonic sector
of a dimensionally reduced string/supergravity theory, and
non-trivially coupled to a negative cosmological constant $\Lambda$
and to a 3-brane of  positive tension $T_3$:
\bea
S=S_{\rm bulk}+ S_{\rm brane}&=&{M_5^3}\int d^5 x \sqrt{|g|}
\left( -R+\frac{1}{2} g^{AB} \partial_A \phi
\partial_B \phi -2 \Lambda e^{\alpha_1 \phi}
\right) \nonumber\\
&-& \frac{T_3}{2} \int d ^4 \xi \sqrt{|\gamma|}\left[
\gamma^{\alpha \beta} \partial_\alpha X^A \partial_\beta
 X^B g_{AB}
e^{\alpha_2 \phi} -2 \right].
\label{21}
\eea
Here $M_5$ is the fundamental mass scale of the five-dimensional bulk
space-time, and the parameters $\alpha_1, \alpha_2$ control the
coupling of the bulk dilaton to $\Lambda$ and to the brane (particular
values of these parameters may simply correspond to the rescaling of
the minimally coupled $\sigma$-model action in the canonical Einstein
frame \cite{28}, but here we allow in general for non-minimal
couplings). The brane action (see for instance \cite{29}) is parametrized
by the coordinates $X^A(\xi)$ describing the embedding of the brane in
the bulk manifold, and by the auxiliary metric tensor $\gamma_{\alpha
\beta}(\xi)$ defined on the four-dimensional world-volume of the
brane, spanned by the coordinates $\xi^\alpha$. Consequently,
$\pa_{\a}  X^A$ is a short-cut notation for $\partial X^A(\xi)/\partial
\xi^\alpha$.

Conventions: Greek
indices run from $0$ to $3$, capital Latin indices from $0$ to $4$,
lower-case Latin indices from $1$ to $3$. For the bulk coordinates we use the
notation $x^A= (t,x^i,z)$. The metric signature is $(+,-,-,-,-)$, and the
curvature tensor is defined by ${R_{MNA}}^B=\partial_M
{\Gamma_{NA}}^B+{\Gamma_{MP}}^B {\Gamma_{NA}}^P - (M
\longleftrightarrow N)$, $R_{NA}={R_{MNA}}^M$.

The variation of the action with respect to $g_{AB}$, $\phi$, $X^A$ and
$\ga_{\a\b}$ gives, respectively, the Einstein equations (in units
such that $M_5^3=1$):
\begin{eqnarray}
&{G^A}_B & =\frac{1}{2}\left(
\partial^A \phi \partial_B \phi -
\frac{1}{2}\delta^A_{B}\partial_C \phi
\partial^C \phi \right) + \Lambda e^{\alpha_1 \phi}\delta^A_B+
\nonumber \\
&&+\frac{T_3}{2} \frac{1}{ \sqrt{|g|}} g_{BC}\int d ^4 \xi
\sqrt{|\gamma|} \delta^5(x-X) \gamma^{\alpha \beta}
\partial_\alpha X^A \partial_\beta
X^C e^{\alpha_2 \phi},
\label{Eq Einstein}
\end{eqnarray}
the dilaton equation:
\begin{equation}
\nabla_M \nabla^M  \phi +2\alpha_1 \Lambda e^{\alpha_1\phi} +\frac{\alpha_2
T_3}{2}\frac{1}{\sqrt{|g|}} \int d ^4 \xi \sqrt{|\gamma|}
\delta^5(x-X) \gamma^{\alpha \beta}
\partial_\alpha X^A
\partial_\beta X^B g_{AB} e^{\alpha_2 \phi}
=0,
\label{Eq dilaton}
\end{equation}
the equation governing the evolution of the brane in
the bulk space-time:
\begin{equation} \partial_\alpha
\left(\sqrt{|\gamma|}\gamma^{\alpha\beta} \partial_\beta X^B
g_{AB} e^{\alpha_2 \phi}\right)=\frac{1}{2}
\sqrt{|\gamma|}\gamma^{\alpha \beta} \partial_\alpha X^B
\partial_\beta X^C \left.
\pa_A  \left(g_{BC} e^{\alpha_2 \phi} \right) \right|_{x=X(\xi)} ,
\label{Eq brane}
\end{equation}
and the induced metric on the brane:
\begin{equation}
\gamma_{\alpha \beta} = \partial_\alpha X^A \partial_\beta X^B
g_{AB} e^{\alpha_2 \phi}
\label{Eq gamma}.
\end{equation}

We now  specialize these equations to the case of a conformally
flat bulk metric, with a warp factor $a$ and a dilaton $\phi$, both of  
which only
depend on the fifth coordinate $z$. Also, we shall look for
$Z_2$-symmetric solutions, describing a flat brane rigidly located at
$z=0$, and we set
\beq
g_{AB}= a^2(z)\eta_{AB}, ~~~~~~ \phi = \phi(z), ~~~~~~
X^A =  \delta^A_\mu\xi^\mu.
\label{26}
\eeq
where $\eta_{AB}$ is the five-dimensional Minkowski metric. The
induced metric thus reduces to
\begin{equation}
\gamma_{\alpha \beta} = \delta^A_\alpha \delta^B_\beta g_{AB}~
e^{\alpha_2 \phi},
\end{equation}
while the brane equations (\ref{Eq brane}) are identically satisfied
 thanks to the $Z_2$ symmetry.

The dynamical equations are obtained from the dilaton equation
(\ref{Eq dilaton}), which becomes
\begin{equation}
3\frac{a'}{a}\phi'+\phi''-2\alpha_1 \Lambda a^2 e^{\alpha_1
\phi}-2\alpha_2
T_3 a e^{2\alpha_2 \phi} \delta(z)=0 ,
\label{Eq reduced dilaton}
\end{equation}
and from the $(\a,\b)$ and $(4,4)$ components of the Einstein equations
(\ref{Eq Einstein}), which give, respectively,
\bea
&&
-3\frac{a''}{a}
=\frac{\phi'^2}{4}+\Lambda a^2 e^{\alpha_1
\phi}+\frac{T_3}{2}a e^{2\alpha_2 \phi} \delta(z),
\label{Eq G00}\\
&&
-6\frac{a'^2}{a^2} = -\frac{\phi'^2}{4}+\Lambda a^2 e^{\alpha_1
\phi}
 \label{Eq G44}
\eea
(a prime denotes differentiation with respect to $z$). The last three
equations are  not
independent:   the dilaton equation, for instance,  can be obtained
by differentiating eqs.  (\ref{Eq G00}), (\ref{Eq G44}), as a consequence
of  the Bianchi identities.

No general solution is known for arbitrary values of
$\alpha_1,\alpha_2, T_3$ and $\La$. However, if we fine-tune these  
parameters by choosing:
\beq
\alpha_1= 4 \alpha_2 \;,~~~ T_3 = 8 \sqrt{\La/\Delta}\;, ~~~  
\alpha_1^2=\Delta+\frac{8}{3}\; ,
\label{alpha rel}
\end{equation}
where the last equation defines $\Delta$,
we recover the
four-dimensional sector of a known, one-parameter family of exact
domain wall solutions \cite{30}, which can be written in an
explicitly $Z_2$-symmetric form.
For $\Delta = -2$ the solution is:
\beq
a(z) = e^{-\frac{k|z|}{3}}, ~~~~~~
\phi(z)=\sqrt{\frac{2}{3}} k |z|,~~~~~~ k^2=-2 \Lambda,\nonumber  \\
\label{213}
\eeq
otherwise ($\Delta \not= -2$):
\beq
a(z) =\left( 1+k |z| \right)^{\frac{2}{3(\Delta+2)}}, ~~~~~~
e^{\phi(z)}=
\left( 1+k |z| \right)^{-\frac{2\alpha_1}{\Delta+2}},~~~~~~
k^2=\frac{(\Delta+2)^2 \Lambda}{\Delta}.
 \label{214}
\eeq

We shall choose $k>0$ so that the $z$ coordinate, transverse to the
brane, may run from $-\infty$ to $+\infty$ (the proper size of the
transverse dimension is finite, however, unless $\Da=-8/3$). In
that case, the solution corresponds to a brane of positive tension,
$T_3>0$, provided $\Da \leq -2$. This range of $\Da$ guarantees a
positive tension and also avoids the presence of naked singularities
\cite{27}. On the other hand, the reality of $\a_1$ requires $\Da \geq
-8/3$. In the rest of this paper we shall thus assume
\beq
k>0, ~~~~~~~~~~~~~~~~~~~-{8\over 3} \leq \Da \leq -2.
\label{215}
\eeq
We may note that in the limit  $\Da = -8/3$ the dilaton decouples and
becomes trivial, and the solution reduces to the well studied pure
AdS$_5$ background , originally introduced to localize gravity
on a 3-brane \cite{9a}.

In the following section we will obtain the canonical equations
governing the evolution of scalar (metric + dilaton) fluctuations around
the above CLP background solutions.

\section{Scalar perturbations}
\setcounter{equation}{0}
\def\theequation{\thesection.\arabic{equation}}

We now perturb to first order the full set of bulk equations
(\ref{Eq Einstein})--(\ref{Eq gamma}), keeping the position of the
brane  fixed, $\da X^A=0$. It is known, indeed, that the brane location
can be consistently assumed to remain unchanged, to the relevant linear
order, when perturbing a background of the type we are considering
\cite{16} (see \cite{21} for a study that includes, instead, a
non-vanishing deformation of the brane). We thus set
\beq
\delta g_{AB}=h_{AB}, ~~~~~~ \delta g^{AB}=-h^{AB}, ~~~~~~
\delta\phi=\chi , ~~~~~~ \da X^A=0,
\label{Pert dil}
\eeq
where the
indices of the perturbed fields are raised and lowered by the
unperturbed metric, and the background fluctuations $h_{AB}, \chi$ are
assumed to be inhomogeneous.

The perturbation of the background equations  (\ref{Eq
Einstein})--(\ref{Eq gamma}) gives, respectively, the linearized
equations for the Einstein tensor:
\begin{eqnarray}
&&
\delta {G^A}_B  =\frac{1}{2} \left(-h^{AC}\partial_C \phi
\partial_B \phi+\partial^A \phi \partial_B \chi+ \partial^A \chi
\partial_B \phi \right)-\frac{1}{4} \delta^A_B \left( 2\partial ^M
\partial_M \chi -h^{MN} \partial_M \phi \partial_N \phi \right)
\nonumber \\
&& + \Lambda \alpha_1 \chi e^{\alpha_1 \phi}
\delta^A_B+\frac{T_3}{2}\frac{1}{\sqrt{|g|}} \int d ^4 \xi
\sqrt{|\gamma|} \delta^5(x-X) \gamma^{\alpha \beta}
\partial_\alpha X^A \partial_\beta
X^C e^{\alpha_2 \phi} \nonumber
\\ &&
\times
\left[ h_{BC}+ g_{BC}
\left( -\frac{1}{2} g^{MN} h_{MN} +\frac{1}{2} \gamma^{\mu \nu}
\delta \gamma_{\mu \nu}+ \alpha_2 \chi \right) \right]
\nonumber
\\ &&
+\frac{T_3}{2}\frac{1}{\sqrt{|g|}}g_{BC}
\int d ^4 \xi \sqrt{|\gamma|} \delta^5(x-X)
\partial_\alpha X^A \partial_\beta
X^C e^{\alpha_2 \phi}\delta \gamma^{\alpha\beta},
\label{32}
\end{eqnarray}
for the dilaton:
\begin{eqnarray}
&&\nabla_M \nabla^M \chi -h^{MN} \nabla_M \nabla_N \phi-g^{MN}
\delta {\Gamma_{MN}^B} \partial_B \phi+ 2 \alpha_1^2 \Lambda \chi
e^{\alpha_1 \phi}
\nonumber \\
&&
+\frac{\alpha_2 T_3}{2}\frac{1}{\sqrt{|g|}}
\int d ^4 \xi \sqrt{|\gamma|} \delta^5(x-X)
\gamma^{\alpha \beta}
\partial_\alpha X^A \partial_\beta
X^B g_{AB} e^{\alpha_2 \phi} \nonumber\\
&&
\times
\left( -\frac{1}{2} g^{MN} h_{MN}
+\frac{1}{2} \gamma^{\mu \nu} \delta \gamma_{\mu \nu}+ \alpha_2
\chi \right)\nonumber
\\ &&
+\frac{\alpha_2 T_3}{2}\frac{1}{\sqrt{|g|}} \int d ^4
\xi \sqrt{|\gamma|} \delta^5(x-X)
\partial_\alpha X^A \partial_\beta
X^B  e^{\alpha_2 \phi}\left( g_{AB}\delta
\gamma^{\alpha\beta}+h_{AB} \gamma^{\alpha\beta} \right)=0,
\end{eqnarray}
for the brane:
\begin{eqnarray}
&&\partial_\alpha \left\{ \sqrt{|\gamma|}
\partial_\beta X^B e^{\alpha_2 \phi} \left[ \delta
\gamma^{\alpha\beta}
 g_{AB} +\gamma^{\alpha\beta}
 h_{AB}+ \gamma^{\alpha\beta}
 g_{AB} \left(\alpha_2 \chi +\frac{1}{2} \gamma^{\mu\nu}  \delta
\gamma_{\mu\nu}
 \right) \right]  \right\}= \nonumber \\
 &&=\frac{1}{2}
\sqrt{|\gamma|} \partial_\alpha X^B
\partial_\beta X^C \left[ \left( \delta\gamma^{\alpha \beta}
+\frac{1}{2} \gamma^{\alpha \beta} \gamma^{\mu \nu}\delta
\gamma_{\mu \nu} \right)\left. \frac{\partial}{\partial x^A}
\right|_{x=X(\xi)} \left( g_{BC} e^{\alpha_2 \phi} \right)
\right. \nonumber \\
&& \left.+ \gamma^{\alpha \beta}\left.
\pa_A \left( h_{BC}
e^{\alpha_2 \phi} +\alpha_2 \chi g_{BC} e^{\alpha_2 \phi}\right)
\right|_{x=X(\xi)}
\right],
\end{eqnarray}
and for the induced metric:
\begin{equation}
\delta \gamma_{\alpha \beta} = \partial_\alpha X^A \partial_\beta
X^B  e^{\alpha_2 \phi} \left( h_{AB} + \alpha_2 \chi g_{AB}
\right).
\label{35}
\end{equation}
Here all geometrical quantities, such as the perturbed connection $\da
\Ga_{AB}\,^C$, the perturbed scalar curvature $\da R= -h^{AB}R_{AB}+
g^{AB}\da R_{AB}$,
and so on, are computed to first order in $h_{AB}$.

By expanding around the background (\ref{26}), it is now easy
to study the propagation  of the spin-2 physical degrees of freedom
on the brane, represented by the transverse and traceless
perturbations $\overline{h}_{ij}$,
\beq
h_{AB}=a^2\delta^i_A \delta^j_B \overline{h}_{ij}, ~~~~~~~~~~~
{\overline{h}^i}_i=0,  ~~~~~~~~~~~~
\nabla^i \overline{h}_{ij}=0.
\label{36}
\eeq
In the linear approximation, the tensor fluctuations $\overline{h}_{ij}$
are decoupled from the scalar and matter fluctuations. We can
consistently set $\chi=0$, and  find that the dilaton and
the brane equations
are trivially satisfied; in addition, the right-hand side of the Einstein
equations (\ref{32}) is identically vanishing, and the linearized Ricci
tensor leads to the well known covariant wave equation for
gravitons:\begin{equation}
\Box_5 \overline{h}_{ij}\equiv \left( {\pa^2\over \pa t^2}-
{\pa^2\over \pa x^{i2}} -{\pa^2\over \pa z^2}-{3 a'\over a}
{\pa\over \pa z}\right)\overline{h}_{ij}=0,
\label{Eq libera}
\end{equation}
where $\Box_5 \equiv \nabla_M \nabla^M$ is the five-dimensional
covariant d'Alembert operator, describing free propagation in the
warped bulk geometry.

In this paper (also in preparation of future cosmological
applications) we are primarily interested in the scalar fluctuations of
the bulk metric, which are coupled to the dilaton fluctuations. Thus, we
shall keep $\da \phi =\chi \not=0$, and we shall expand around the
background (\ref{26})  in the so-called
``generalized longitudinal gauge" \cite{16}, which extends the
longitudinal gauge of standard cosmology \cite{31} to the brane-world
scenario. As discussed in \cite{16}, in five dimensions there are four  
independent degrees of freedom for the scalar metric
fluctuations: in the generalized longitudinal gauge they are
described by the four variables $\{ \varphi, \psi, \Gamma, W\}$, defined
by \bea
&&
h_{00}= 2 \varphi a^2, ~~~~~~~~~~~~~~~
h_{ij}=2\psi a^2 \da_{ij}, \nonumber\\
&&
h_{44}= 2 \Ga a^2, ~~~~~~~~~~~~~~~
h_{04}=-W a^2.
\label{38}
\eea
(Off-diagonal metric fluctuations have been taken into account also in
a recent study of linearized gravity in a brane-world background
\cite{32}: in that case, however, there are no scalar sources in the bulk,
and no long-range scalar interactions).

By inserting the explicit form of the background (\ref{26}) and of the
metric fluctuations (\ref{38}) into the perturbed equations
(\ref{32})--(\ref{35}), we  obtain the full set of constraints
and dynamical equations governing the linearized evolution of the five
scalar variables $\{ \varphi, \psi, \Gamma, W,\chi\}$. Let us give them
in components, starting from the Einstein equations, and using eq.
(\ref{35}) for the perturbations of the induced metric.

 Equation $(0,0)$ gives:
\begin{eqnarray}
&&2\nabla^2 \psi+ 3\psi'' +\nabla^2\Gamma
+9\frac{a'}{a}\psi' -3\frac{a'}{a}\Gamma' -\frac{\phi'}{2}\chi'
-6\frac{a''}{a}\Gamma -\frac{{\phi'}^2}{2}\Gamma \nonumber \\
&&-a^2 e^{\alpha_1 \phi} \Lambda \alpha_1 \chi - \frac{1}{2} a
e^{2\alpha_2\phi} T_3 \left( \Gamma+ 2\alpha_2 \chi \right)
\delta(z)=0. \label{Eq00}
\end{eqnarray}

Equation  $(i,i)$ gives:
\begin{eqnarray}
&& -\nabla^2 \varphi -\varphi'' -2\ddot \psi +\nabla^2 \psi
+2\psi'' -\ddot \Gamma +\nabla^2 \Gamma \nonumber \\ &&
-3\frac{a'}{a} \varphi' -\dot W'
-3\frac{a'}{a}\dot W +6
\frac{a'}{a} \psi'
-3\frac{a'}{a}\Gamma'
-\frac{\phi'}{2}\chi'
-6\frac{a''}{a}\Gamma-\frac{{\phi'}^2}{2}\Gamma \nonumber \\ &&
-a^2 e^{\alpha_1 \phi} \Lambda \alpha_1 \chi - \frac{1}{2} a
e^{2\alpha_2\phi} T_3 \left( \Gamma+ 2\alpha_2 \chi \right)
\delta(z)=0. \label{Eqii}
\end{eqnarray}

Equation  $(i,j)$, with $i \neq j$, gives:
\begin{equation}
\partial_i\partial_j \left( \varphi -\psi -\Gamma \right)=0.
\label{Eqij}
\end{equation}

Equation  $(4,4)$ gives:
\begin{eqnarray}
&& -\nabla^2 \varphi -3\ddot\psi +2\nabla^2\psi
-3\frac{a'}{a}\varphi' -3\frac{a'}{a}\dot W +9\frac{a'}{a} \psi'
+\frac{{\phi'}}{2}\chi' \nonumber \\ &&-12\frac{a'^2}{a^2}
\Gamma+\frac{{\phi'}^2}{2}\Gamma-a^2 e^{\alpha_1 \phi} \Lambda
\alpha_1 \chi =0. \label{Eq44}
\end{eqnarray}

Equation  $(i,0)$ gives:
\begin{equation}
\partial_i \left(\frac{W'}{2} +\frac{3}{2}\frac{a'}{a}W +2 \dot \psi +\dot
\Gamma \right) =0. \label{Eq0i}
\end{equation}

Equation  $(4,0)$ gives:
\begin{equation}
\frac{1}{2}\nabla^2 W -3\dot \psi' +3\frac{a'}{a} \dot \Gamma
+\frac{\phi'}{2} \dot \chi=0. \label{Eq40}
\end{equation}

Equation  $(4,i)$ gives:
\begin{equation}
\partial_i \left( \frac{\dot W}{2} +\varphi' -2\psi'
+3\frac{a'}{a} \Gamma +\frac{\phi'}{2} \chi\right)=0. \label{Eq4i}
\end{equation}

The dilaton equation gives:
\begin{eqnarray}
&& \Box_5 \chi  - \phi' \varphi' -\phi' \dot W +3 \phi' \psi' -\phi'
\Gamma'  -6 \frac{a'\phi'}{a}\Gamma -2\phi''\Gamma \nonumber \\
&& +2a^2 e^{\alpha_1 \phi} \Lambda \alpha_1^2 \chi +2 a
e^{2\alpha_2\phi} T_3 \alpha_2 \left( \Gamma+ 2\alpha_2 \chi \right)
\delta(z)=0 \label{Eq prechi}
\end{eqnarray}
(the dots denote differentiation with respect to  Minkowski time
on the brane). Finally, the brane perturbation equation gives a
constraint at $z=0$ which is always satisfied because of the $Z_2$
symmetry.
A similar set of equations
was already derived in \cite{16}. We have no contributions
from the time derivatives of the background, since our background is
static.

In the absence of bulk sources with anisotropic stresses, we can now
eliminate $\varphi$ from eq. ($i \not=j$), thus reducing to four scalar
degrees of freedom, by setting:
\begin{equation}
\varphi=\psi+\Gamma. \label{318}
\end{equation}
As a consequence, we immediately find that the variable $W$ decouples
from the other fluctuations. The combination of the time derivative of
eq. ($4,i$) with the $z$-derivative of eq. ($i,0$) and with eq. ($4,0$)
leads in fact to the equation
\begin{equation}
\Box_5 W=3\left( \frac{a''}{a}-\frac{a'^2}{a^2} \right)W.
\label{319}
\end{equation}
Thus $W$ is decoupled but, because of the non-trivial self-interactions,
it does not freely propagate in the background geometry like the
graviton, eq. (\ref{Eq libera}).

In order to discuss the dynamics of the remaining variables $\psi,
\Gamma$ and $\chi$, it is now convenient to recombine their
differential equations in an explicitly covariant way, to obtain a
canonical evolution equation. To this aim, we can use eq. (\ref{318})
and eq. ($4,i$) to eliminate $\varphi$ and $\dot W$ in eqs. ($i,i$), ($4,4$)
and in the dilaton equation (\ref{Eq prechi}). We then combine the
simplified version of eqs. ($4,4$), ($i,i$) with eq. ($0,0$), and  obtain
the following system of coupled equations, where the source terms
 depend only on $\Gamma$ and $\chi$:
\begin{eqnarray}
\Box_5 \psi
&=&-2\frac{a''}{a}\Gamma+2\frac{a'^2}{a^2}\Gamma
+\frac{a'\phi'}{a}\chi -\frac{2}{3}a^2 e^{\alpha_1 \phi} \Lambda
\alpha_1 \chi \nonumber\\
&- &\frac{1}{6} a e^{2\alpha_2\phi} T_3 \left( \Gamma+
2\alpha_2 \chi \right)
\delta(z), \label{Eq psi} \\
 \Box_5 \Gamma&= &-2\frac{a''}{a}\Gamma
+8\frac{a'^2}{a^2}\Gamma -\phi'^2\Gamma +\phi''\chi
+\frac{a'\phi'}{a}\chi -\frac{2}{3}a^2 e^{\alpha_1 \phi} \Lambda
\alpha_1 \chi \nonumber\\
&-&
\frac{2}{3} a e^{2\alpha_2\phi} T_3 \left( \Gamma+
2\alpha_2 \chi \right)
\delta(z), \label{Eq Gamma} \\
\Box_5 \chi &=&2\phi''\Gamma -\phi'^2\chi -2a^2
e^{\alpha_1 \phi} \Lambda \alpha_1^2 \chi \nonumber\\
&-& 2 a e^{2\alpha_2\phi}
T_3 \alpha_2 \left( \Gamma+ 2\alpha_2 \chi \right) \delta(z).
\label{Eq chi}
\end{eqnarray}

This system can be diagonalized by introducing the fields
\beq
\omega_1=2\psi+ \Gamma, ~~~~~
\omega_2=6\alpha_2 \Gamma +\chi, ~~~~~
\omega_3=\Gamma- 2\alpha_2 \chi,
\label{323}
\eeq
relations that can be inverted as:
\begin{eqnarray}
&&\psi=\frac{\omega_1}{2}-\frac{2\alpha_2 \omega_2
+\omega_3}{2\left(1+12\alpha_2^2\right)}, ~~~~~~~
 \Gamma=\frac{2\alpha_2 \omega_2
+\omega_3}{\left(1+12\alpha_2^2\right)}\nonumber\\
&& \chi=\frac{\omega_2-6\alpha_2
\omega_3}{\left(1+12\alpha_2^2\right)}.
\label{324}
\end{eqnarray}
In terms of these new variables, the perturbation equations  (\ref{Eq
psi})--(\ref{Eq chi}) reduce to
\begin{eqnarray}
  \Box_5 \omega_1&=&0, \label{325}\\
\Box_5 \omega_2&=&0 , \label{326}\\
 \Box_5 \omega_3&=&
\left[-\frac{\phi''}{2\alpha_2}
-\phi'^2-\frac{a'\phi'}{2\alpha_2a} +
\left(\frac{\alpha_1}{3\alpha_2}-2\alpha_1^2 \right)a^2
e^{\alpha_1 \phi} \Lambda\right]\om_3
\nonumber\\
&+&
  \left[\left( \frac{2}{3}-4\alpha_2^2
\right) a e^{2\alpha_2\phi} T \delta (z)\right]\omega_3.
\label{327}
\end{eqnarray}

Together with eq. (\ref{319}), and the constraints  
(\ref{Eq0i})--(\ref{Eq4i}),  the above decoupled equations  
describe the
complete evolution of the scalar (metric + dilaton) fluctuations in the CLP
brane-world background (\ref{213}), (\ref{214}). Two variables
($\om_1, \om_2$) are (covariantly) free on the background like the
graviton, while the other two variables  ($\om_3, W$) have non-trivial
self-interactions.

In all cases, we can introduce the corresponding ``canonical
variables" $\ha W$, $\ha \om_i$ ($i=1,2,3$), which have
canonically normalized   kinetic
terms
\cite{31} in the action, simply by absorbing the
geometric warp factor as follows:
\begin{equation}
W= \ha W a^{-3/2}, ~~~~~~~~~~~~~~~
\om_i= \ha \om_i a^{-3/2}.
\label{328}
\end{equation}
The variables $\ha W$, $\ha \om_i$ are required for a correct
normalization of the scalar perturbations to a quantum fluctuation
spectrum, as they satisfy canonical Poisson (or commutation) brackets.
When the general solution is written as a  superposition of free,
factorized plane-waves modes
on the brane,
\beq
\ha W= \Psi_w(z)  e^{-ip_\mu x^\mu}, ~~~~~~~~~~~
\ha \om_i= \Psi_i(z)  e^{-ip_\mu x^\mu},
\label{329}
\eeq
they define the inner product of states with measure $dz$  \cite{17}, as in
conventional one-dimensional quantum mechanics, $\int dz
|\Psi(z)|^2 $.

The allowed mass spectrum of $m^2=\eta^{\mu\nu} p_\mu p_\nu$, for
the scalar fluctuations on the brane, can then be obtained by solving an
eigenvalue problem in the Hilbert space $L^2(R)$ for the canonical
variables $\Psi_w,\Psi_i$, satisfying a Schr\"odinger-like equation in $z$,
which is obtained from the equations  (\ref{319}),
(\ref{325})--(\ref{327})  for $W$ and
$\om_i$, and which can be written in the conventional form as:
\beq
\Psi_w''+\left(m^2- {\xi_w''\over \xi_w}\right)\Psi_w=0, ~~~~~~~~~~~~
\Psi_i''+\left(m^2- {\xi_i''\over \xi_i}\right)\Psi_i=0.
\label{330}
\eeq
Here, by analogy with cosmological perturbation theory \cite{31}, we
have introduced four ``pump fields" $\xi_w,\xi_i$, defined  as follows:\bea
&&
\xi_w=a^{\b_w}, ~~~~~~~~~~~~ \xi_i=a^{\b_i}, \nonumber\\
&&
\b_w=-{3\over 2}, ~~~~~~~~~ \b_1=\b_2 ={3\over 2}, ~~~~~~~~~
\b_3=-{1\over 2}(1+3 \a_1^2)=-{3\over 2}(\Da+3).
\label{332}
\eea

The effective potential generated by the
derivatives of the pump fields depends on $\b_w,\b_i$, and contains in
general a smooth part, peaked at $z=0$, plus a positive or negative
$\da$-function contribution at the origin. We may have, in principle, not
only volcano-like potentials, which correspond to the free covariant
d'Alembert equation with $\b=3/2$ \cite{27} (and which are known to
localize  gravity \cite{9a,17}), but also potentials that are positive
everywhere and admit no bound states. The possible localization of
scalar interactions on the 3-brane, for the given background and
perturbation equations, will be discussed in the next section.

\section{Localization of the massless modes}
\setcounter{equation}{0}
\def\theequation{\thesection.\arabic{equation}}

The general solutions of the canonical perturbation equations (\ref{330})
are labelled by the mass eigenvalue $m$, by their parity with respect to
$z$-reflections, and by the parameters $\b_w,\b_i$, which depend on
the type of perturbation. The massless scalar modes corresponding to
bound states of the effective potential \cite{9a,17,27} will describe
long-range scalar interactions confined to the four-dimensional brane.
Concerning $z$-reflections, we shall follow  the
perturbative formalism developed in \cite{16} consistently,   
restricting ourselves to
a perturbed background that is still $Z_2$-symmetric, namely to
$Z_2$-odd solutions for $W$, and to $Z_2$-even solutions for $\om_i$.
To describe a bound state, we shall further restrict such solutions to
those with a normalizable canonical variable  w.r.t. the measure $dz$,
namely to $\Psi(z)~ \in ~ L^2(R)$. Among the acceptable solutions, we
shall finally select those satisfying the constraints  
(\ref{Eq0i})--(\ref{Eq4i}) derived in the previous
section. The above set of conditions will determine the class of
brane-world backgrounds allowing the four-dimensional localization of
long-range scalar interactions.

For  $m=0$ the  exact solution of the canonical equations (\ref{330}),
for a generic pump field $\xi(z)$, can be separated into an even and
an odd part, $\Psi^+$ and $\Psi^-$, as follows:
\beq
\Psi_0^+(z)=c_+\xi(z), ~~~~~~~~~~~
\Psi_0^-(z)= c_-\xi(z)\int dz'
\xi^{-2}(z'),
\label{41}
\eeq
where $c_+$ and $c_-$ are integration constants. In order to
parametrize the solutions for different values of $\Da$ and $\b$ (and
also in view of  subsequent applications to the massive mode
solutions), it is now convenient to introduce the two indices $\nu$ and
$\nu_0$, defined by
\beq
 \nu_0=\frac{\Delta}{2(\Delta+2)}, ~~~~~~~~~~~~
\nu=\frac{1}{2}-\frac{2\beta}{3(\Delta+2)}.
\label{42}
\eeq
For the three possible values of $\beta_w, \b_i$ (see eq. (\ref{332})),
they are related by
\begin{eqnarray}
&&\nu=1-\nu_0, \hspace{2.25cm}{\b=\b_w=-{3\over 2}} , \nonumber\\
&&\nu=\nu_0, \hspace{2.85cm}{\b=\b_{1,2}={3\over 2}}, \nonumber\\
&&\nu=2-\nu_0, \hspace{2.25cm}{\b=\b_3=-{3\over 2}(\Da+3) }.
\label{43}
\end{eqnarray}
We
recall that, in the class of backgrounds we are considering, the
parameter $\Da$ is constrained in the range $-8/3\leq \Da \leq
-2$ (because of the conditions of positive tension, absence of naked
singularities and reality of the dilaton couplings). As a consequence, the
index $\nu_0$ can range from $2$ to $+\infty$.

Let us  first  discuss the limiting case $\nu=\nu_0=\infty$,
corresponding to $\Da=-2$.  For a
generic  pump field, with exponent $\beta$, the massless solutions are:
\beq
\Psi_{0,\infty}^+(z) =  c^+_{0,\infty}e^{-\frac{\beta}{3} k|z|}, ~~~~~~~~~~
\Psi_{0,\infty}^-(z)= c^-_{0,\infty}{\rm sgn}\{z\}
\left(e^{\frac{\beta}{3} k|z|}-e^{-\frac{\beta}{3}
k|z|}\right).
\label{44}
\eeq
In our background $k>0$, so that the fluctuations $W$ and $\om_3$,
both with $\b=-3/2$, are not normalizable.  The even solutions of
the free d'Alembert equation, with $\beta=3/2$, are instead
normalizable, so we have acceptable solutions  for $\omega_1$ and
$\omega_2$. However, from the constraint ($i,0$) of eq. (\ref{Eq0i}),
$\om_1$ is forced to vanish when $W=0$, unless the fluctuations are
static, $\dot \om_1=0$.
All the other constraints
(using also the background equations) are instead identically satisfied
by $\om_2$.
It follows that, for $\Da=-2$,
there are two independent massless modes localized on the brane: one,  
$\omega_2$,
is propagating and the other, $\om_1$,  is static.

The same is true for the case $\Da<-2$, i.e. finite $\nu_0$. In that case
 the massless solutions can be written in the form
\begin{eqnarray}
&& \Psi_{0,\nu}^+(z)=  c^+_{0,\nu} \left(1+k|z|
\right)^{\frac{1}{2}-\nu },\nonumber
\\ &&
\Psi_{0,\nu}^-(z)=  c^-_{0,\nu}  {\rm sgn}\{z\}\left[ \left(1+k|z|
\right)^{\frac{1}{2}+\nu } -\left(1+k|z| \right)^{\frac{1}{2}-\nu
}\right]. \label{45}
\end{eqnarray}
The solutions for $W$ and $\om_3$ always correspond to $\nu<0$, and
are not normalizable. The even d'Alembert modes, with $\nu=\nu_0$,
are normalizable for $\nu>1$, i.e. $\Da >-4$, so that $\om_1$ and
$\om_2$ are always acceptable in our class of backgrounds, for which
$\Da \geq -8/3$. Again, however, the constraint (\ref{Eq0i}) implies
$\om_1=0$, unless  $\dot \om_1=0$, while the other constraints are
identically satisfied by $\om_2$, so we are left with  non-trivial
massless solutions only for  a propagating fluctuation, $\om_2$,  and for
a static one, $\om_1$.

We may thus conclude that all backgrounds of the class defined by the
conditions (\ref{215}) localize on the brane not only the
massless spin-2 degrees of freedom \cite{27} (long-range tensor
interactions), but also  one propagating massless scalar
degree of freedom ($\om_2$),  corresponding to a long-range scalar
interaction generated by the dilaton field.
In the longitudinal gauge, the
canonical representation $\om_2$ of such a scalar interaction is
associated not
only   with the dilaton fluctuation $\chi$, but also with the
four-dimensional ``Bardeen potential" $\psi$, and with the ``breathing
mode" $\Ga$ of the dimension orthogonal to the brane (see eqs.
(\ref{324})). In addition, we have a second independent massless
degree of freedom localized on the brane ($\om_1$), which is not
propagating ($\dot \om_1=0$), but is essential  to
reproduce the standard long-range gravitational interaction in the static
limit, as we shall discuss in Section 6.

The localization of the scalar
interactions does not impose any further constraints on the
background, besides those of eq. (\ref{215}).  Also,  in
the limiting case of a
pure AdS$_5$ solution  ($\Da=-8/3$, $\nu_0=2$, $\a_1=0=\a_2$) the
dilaton background disappears, and the dilaton fluctuation
$\chi=\om_2$ decouples from the others. The only (static) contribution
to the scalar sector of metric fluctuations comes from $\om_1$, which
generates the long-range Newton potential $\varphi=\psi$ on the brane
(see Section 6). However, even in the case $\Da=-8/3$, the propagating
dilaton flucutuation  $\om_2$ is non-vanishing, and remains there to
describe a (possibly dangerous) long-range scalar interaction. By  
contrast, no propagating
massless
scalar modes appear in the ``pure-gravity" models without bulk scalar
fields in the action, as the AdS$_5$ brane-world scenario discussed in
\cite{9a}.

\section{The massive mode spectrum}
\setcounter{equation}{0}
\def\theequation{\thesection.\arabic{equation}}

The massive part of the spectrum of the canonical equations (\ref{330})
is not localized on the brane; it  may induce higher-dimensional (and
short-range) corrections to the long-range scalar forces (just as
in the case of the pure tensor part of the gravitational interaction
\cite{9a,17,27}). Such corrections thus provide physical effects from  
the fifth
 dimension on our brane.

In order to determine the  massive modes able to survive the
constraints of Section 3, and to evaluate the corresponding short-range
corrections, we shall first present the exact solutions of the massive
canonical equations  (\ref{330}), distinguishing, as before, the even and
odd parts under $z$-reflections. For $m\not=0$,  $\Da=-2$ (i.e.
$\nu_0=\infty$), and a generic pump field with power $\b$, such
solutions can be written as follows
\begin{eqnarray}
&& \Psi^+_{m,\infty}(z)=\frac{1}{\sqrt{\pi m q }} \left( q \cos
q|z| -\frac{\beta }{3} k \sin q|z| \right),\nonumber \\
 &&
\Psi^-_{m,\infty}(z)=-\sqrt{\frac{m}{\pi  q } }\sin q z  ,
\label{51}
\eea
where
\beq
q=
\left(m^2-\frac{k^2}{4}\right)^{1/2}.
\label{52}
\eeq
For $\Da<-2$, i.e.  finite $\nu_0$, the solution can be written as a
combination of first-  and second-kind Bessel functions $J_\nu$ and
$Y_\nu$ \cite{33}, of index $\nu$ given by eq. (\ref{42}):
 \begin{eqnarray}
&& \Psi^+_{m,\nu}(z)=c_{m,\nu-1} \sqrt{1+k|z|} \left[ Y_{\nu-
1}\left(\frac{m}{k} \right) J_{\nu }\left(y
\right)-
J_{\nu
-1}\left(\frac{m}{k} \right) Y_{\nu }\left(y
\right)\right] , \nonumber\\
&&\Psi^-_{m,\nu}(z)=c_{m,\nu} ~{\rm
sgn}\{z\}\sqrt{1+k|z|} \left[ Y_{\nu}\left(\frac{m}{k} \right)
J_{\nu }\left(y \right)- J_{\nu }\left(\frac{m}{k} \right) Y_{\nu
}\left(y\right)\right] ,
\label{53}
\eea
where
\beq
c_{m,\nu}= \sqrt{\frac{m}{2k}}  \left[ J_{\nu}^2
\left(\frac{m}{k} \right) + Y_{\nu }^2\left(\frac{m}{k} \right)
\right]^{-\frac{1}{2}}, ~~~~~~~~y= \frac{m}{k}(1+k|z|).
\label{54}
\eeq
Note that in the above equations we have adopted the $\da$-function
normalization of the continuum modes, as for plane waves in
one-dimensional quantum mechanics. As a consequence, $\Psi_m$ is
dimensionless (unlike in \cite{27}, where a different normalization
is adopted).

It is important to note that modes with
 negative squared mass (tachyons) are not
included in the spectrum, as they would not correspond to a
normalizable canonical variable ($\Psi$ would blow up in $z$). This can
be regarded as a direct check of the stability of the given class of CLP
backgrounds against scalar perturbations, since tachyonic modes would
also blow up in time, and would destroy the assumed homogeneity of
the four-dimensional brane. Another consequence of the normalization
condition is the mass gap ($m^2>k^2/4$, see eq. (\ref{52})) between
the localized massless mode and the massive corrections, in the
limiting background with $\nu_0=\infty$ (already noticed in \cite{27}
for the case of pure tensor interactions).

We shall now impose the constraints (\ref{Eq0i})--(\ref{Eq4i})
following from the perturbed background equations. It is convenient to
introduce the four amplitudes $A_i, A_w$, defined by
 \begin{eqnarray}
&& \omega_1=\frac{a^{-3/2} A_1}{c_{m,\nu_0-1}}
\Psi_{m,\nu_0}^+(z)e^{-ip_\mu x^\mu} , ~~~~~~~~~~~~~
\omega_2=\frac{a^{-3/2} A_2}{c_{m,\nu_0-1}}
\Psi_{m,\nu_0}^+(z)e^{-ip_\mu x^\mu}\nonumber \\
 && \omega_3=\frac{a^{-3/2} A_3}{c_{m,1-\nu_0}}
\Psi_{m,2-\nu_0}^+(z)e^{-ip_\mu x^\mu},  ~~~~~~~
W=\frac{a^{-3/2} A_w}{c_{m,1-\nu_0}}
\Psi_{m,1-\nu_0}^-(z)e^{-ip_\mu x^\mu},
\end{eqnarray}
where $\Psi^{\pm}$ are the above normalized solutions.
Plugging this ansatz into the constraint equations
(\ref{Eq0i})--(\ref{Eq4i}), we find that, in contrast with the massless
case, none of the four scalar fluctuations is forced to vanish. However,
only two amplitudes are independent.
By taking, for
instance,  $\om_2$ and $\om_3$ as independent variables, we can
indeed express $A_1$ and $A_w$ in terms of $A_2$ and $A_3$, for all
values of $\nu_0$, in such a way that all the constraints are identically
satisfied. For a generic mode of mass $m$ and momentum $p$,
we find, in particular,
\bea
&&
A_1=\frac{m^2}{9m^2+6p^2}\frac{\sqrt{2\nu_0-1}}{\nu_0-1}\left(
\sqrt{3\left(\nu_0-2 \right)} A_2 +3 \sqrt{2\nu_0-1} A_3 \right),
\nonumber\\
&&
A_w=\frac{2 i m \sqrt{m^2+p^2}}
{9m^2+6p^2}\frac{\sqrt{2\nu_0-1}}{\nu_0-1}\left( \sqrt{3\left(\nu_0-2
\right)} A_2 +3 \sqrt{2\nu_0-1} A_3 \right),
\label{57}
\eea
which also hold when $\nu_0 \rightarrow\infty$. Note that a single  
combination of $A_2$ and $A_3$ determines both $A_1$ and $A_w$.

For such backgrounds we  thus have four types of higher-dimensional
contributions to the scalar interactions on the brane, arising from the
massive spectrum of $\om_i$ and $W$. The exchange of such massive
modes generates corrections to the four-dimensional scalar forces.
The corrections are in general different from those of tensor
interactions, arising from the massive spectrum of a variable
satisfying the free
d'Alembert equation, like $\om_2$. In the weak field limit, however,
the leading-order  contributions to the non-relativistic potential
generated by  a static scalar sources have the same qualitative
behaviour as in the case of tensor interactions, as will be illustrated in
the next section.

\section{Static limit and leading-order corrections}
\setcounter{equation}{0}
\def\theequation{\thesection.\arabic{equation}}

To make contact with previous works, and for future phenomenological
applications, we shall now compute the effective scalar-tensor
interaction induced on the brane, in the weak field limit, by a static and
point-like source of mass $M$ and dilatonic charge $Q$.

In our longitudinal gauge (\ref{36}), (\ref{38}), in which the
decomposition of the metric fluctuations is based on the $O(3)$
symmetry of the spatial hypersurfaces of the brane, the energy
density of a point-like particle only contributes to the scalar part of
the perturbed matter stress tensor (with $T_{00}$ as the only
non-vanishing component), and provides a $\delta$-function source to the
$(0,0)$ scalar perturbation equation (\ref{Eq00}). Similarly, the charge
$Q$ acts as a point-like source in the dilaton perturbation equation
(\ref{Eq prechi}).

As a  consequence, we obtain three $\delta$-function
sources in the equations for the three $\om_i$ fluctuations,
$S_i \da^3(x-x')\da(z)$, with three scalar charges $S_i$, which are
``mixtures"  of $M$ and $Q$, while no source term is obtained in the
static limit for the $W$ fluctuation (in agreement with the fact that
$W=0$, in the static limit, according to the constraint $(i,0)$).  The
effective sources $S_i$ for the massless and massive $\om_i$
fluctuations are defined by the combination of eqs. $(4,4), (i,i), (0,0)$,
and by the dilaton equation (\ref{Eq prechi}), as follows:
\beq
S_1= M, ~~~~~~~~
S_2=Q+2\a_2 M, ~~~~~~~~
S_3= {M\over 3} -2 \a_2 Q.
\label{61}
\eeq

The exact static solutions of eqs. (\ref{325})--(\ref{327}), with the
above point-like sources, can be easily obtained using the static limit
of the retarded Green function evaluated on the brane ($z=0$), i.e.
\beq
\om_i(\nu,x,x')= -S_i G_i (\nu,x,x'),
\label{62}
\eeq
where
\beq
G_i (\nu,\vec x, \vec x', z=z'=0)= \int {d^3p\over (2 \pi)^3} e^{i \vec p
\cdot (\vec x -\vec x')}\left\{{\left[\psi^+_{0,\nu}(0)\right]^2\over p^2}
+\int_{m_0}^\infty dm {\left[\psi^+_{m,\nu}(0)\right]^2\over p^2+m^2}
\right\}
 \label{63}
\eeq
(see also \cite{27,27a,34}). Here $\nu=\nu_0$ for $i= 1, 2$, while
$\nu=2-\nu_0$ for $i = 3$, and $\psi^+_{0,\nu}(z), \psi^+_{m,\nu}(z)$
(including the case $\nu =\infty$) are the exact solutions
(\ref{44}), (\ref{45}), (\ref{51}), (\ref{53}), obtained in the previous  
sections. The
first term in the integrand corresponds to the long-range forces
generated by the massless modes, the second term to the ``short-range"
corrections due to the massive modes, and $m_0$ is the lower bound for
the massive spectrum ($m_0=k/2$ if $\nu_0=\infty$, while $m_o=0$ if
$\nu_0<\infty$, see eqs. (\ref{51}),  (\ref{52})).

We should note that in the $\om_1, \om_2$ case we have to include both
the massless and massive contributions, while in the $\om_3$ case
only the massive ones survive (indeed,  we recall
 that for $m=0$ the $\om_3$
modes are not normalizable, and that the massless sector of $\om_1$,
in the static limit, is not eliminated by the constraint $(i,0)$, see
Section 4). We also note that the amplitude of the massless
solutions (\ref{44}), (\ref{45}) has to be fixed by the correct
normalization, i.e.
\beq
c^+_{0,\infty}= \left(k\b/3\right)^{1/2}, ~~~~~~~~~~~~
c^+_{0,\nu}=\left[k (\nu-1)\right]^{1/2}.
\label{64}
\eeq

Let us start with $\om_1$, for which $\nu=\nu_0$, and with the limiting
background $\Da=-2$, i.e. $\nu=\infty$. By setting $\b=3/2$, and using
eqs. (\ref{44}) and (\ref{51}) for $\psi^+_{0,\infty}$ and
$\psi^+_{m,\infty}$, respectively, we obtain
\beq
\om_1(\nu=\infty) =-S_1 {k\over 8\pi r} \left[1+{2\over k\pi}
\int_{k/2}^\infty {dm\over m} \left(m^2-{k^2\over 4}\right)^{1/2}
e^{-mr}\right],
\label{65}
\eeq
where $r= |\vec x-\vec x'|$. The same integral had already been
obtained in \cite{27} when discussing the localization of tensor
fluctuations, and the associated leading-order corrections (in the  
large-distance limit) are known to be of the form $e^{-kr/2}  
(kr)^{-3/2}$.

In the case $\Da<-2$, i.e. $\nu=\nu_o <\infty$, we shall use instead eqs.
(\ref{45}) and (\ref{53}) for $\psi^+_{0,\nu_0}$ and
$\psi^+_{m,\nu_0}$, respectively. Also, we shall estimate the
contribution of the massive modes by expanding the Bessel functions in
the small-$m$ regime  (which is relevant at the
long distances typical of the weak field limit). The  small
argument limit  \cite{33} of $J_\nu$, $Y_\nu$  then gives
\beq
\left[\psi^+_{m,\nu_0}(0)\right]^2 ={1\over \Ga^2 (\nu_0-1)}
\left(m\over 2k\right)^{2\nu_0-3},
\label{66}
\eeq
and we are led to
\beq
\om_1(\nu<\infty) =-S_1 {k(\nu_0-1)\over 4\pi r} \left[1+
{\Ga (\nu_0-1/2)\over \sqrt \pi \Ga (\nu_0)}\left(1\over
kr\right)^{2\nu_0-2} \right]
\label{67}
\eeq
(here $\Ga$ obviously represents the Euler function, not to be confused
with the  $g_{44}$ component of the metric fluctuations).

Exactly the same results (\ref{65}),  (\ref{67}) are obtained for
$\om_2$, which satisfies the same free d'Alembert equation as
$\om_1$, with the only difference that $S_1$ has to be replaced by
$S_2$.

Let us then compute $\om_3$, for which $\nu=2-\nu_0$, and which has
only the massive mode contribution to the Green function. For $\Da=-2$,
i.e $\nu=\infty$, we shall use eq. (\ref{51}) for $\psi^+_{m,\infty}$. At
$z=0$ the solution however is $\b$-independent, and we obtain the
(massive part) of the
 result already given in eq. (\ref{65}) (with $S_1$ replaced by
$S_3$).

In the case $\Da<-2$,
i.e $\nu<\infty$, we shall use  $\psi^+_{m,2-\nu_0}$ from eq.
(\ref{53}). In the large-distance (small-$m$) regime, however, the
contribution at  $z=0$ to the Green function is exactly the same as that
of eq. (\ref{66}), so that we obtain
\beq
\om_3(\nu<\infty) =-S_3 {k(\nu_0-1)\over 4\pi r}
{\Ga (\nu_0-1/2)\over \sqrt \pi \Ga (\nu_0)}\left(1\over
kr\right)^{2\nu_0-2} .
\label{68}
\eeq

We are now ready to estimate, in the static limit, the dilaton and
metric fluctuations $\chi, \Ga, \psi, \varphi$. We shall explicitly
consider the case $\nu_0<\infty$, for an easy comparison with previous
results relative to an AdS$_5$ background, for which $\nu_0=2$.
Defining
\beq
A_{\nu_0}={k(\nu_0-1)\over 4\pi}, ~~~~~~~~~~~~
B_{\nu_0}={\Ga (\nu_0-1/2)\over \sqrt \pi \Ga (\nu_0)},
\label{69}
\eeq
we first rewrite the $\om_i$ solutions in compact form as:
\bea
&&
\om_1=-{S_1  A_{\nu_0}\over r} \left[1+ B_{\nu_0}\left(1\over
kr\right)^{2\nu_0-2}\right], \nonumber\\
&&
\om_2=-{S_2  A_{\nu_0}\over r} \left[1+ B_{\nu_0}\left(1\over
kr\right)^{2\nu_0-2}\right], \nonumber\\
&&
\om_3=-{S_3  A_{\nu_0}\over r} B_{\nu_0}\left(1\over
kr\right)^{2\nu_0-2} .
\label{610}
\eea
We may note, as a check of our previous computations, that the
relative amplitude of the massive corrections, in the weak, static
limit, is controlled by the correspondig scalar charges, which satisfy  
the
relation (from eq. (\ref{61})):
\beq
S_3+2 \a_2S_2= \left({1\over 3} + 4\a_2^2\right)S_1.
\label{611}
\eeq
By eliminating $\a_2$  in favour of $\Da$ according to eq. (\ref{alpha rel}),
and $\Da$ in terms of $\nu_0$, according to eq. (\ref{42}), we  obtain
\beq
S_1={1\over \nu_0-1} \left[
{1\over 3} \sqrt{3 (\nu_0-2)(2\nu_0-1)}S_2 +(2 \nu_0-1) S_3\right],
\label{612}
\eeq
which exactly reproduces, in the static limit $m^2+p^2 \ra 0$, the
general relation (\ref{57}) between the massive amplitudes.

It is convenient, at this point, to  explicitly introduce the
four-dimensional gravitational constant $M_p^{-2}$, by noting that
\beq
A_{\nu_0}={k(\nu_0-1)\over 4\pi}={1\over 4\pi}
\left[\psi^+_{0,\nu_0}(0)\right]^2={1\over 4\pi}
\left[\int dz~ a^3(z)\right]^{-1}.
\label{613}
\eeq
The above normalization integral, when expressed in terms of a new
bulk coordinate $y$, with $dy= a(z)dz$, represents the warped
extra-dimensional volume that controls the ratio between the four-
and five-dimensional gravitational constants \cite{8,9a}, i.e.
\beq
M_p^2= M_5^3 \int dy ~a^2(y)= M_5^3 \int dz~a^3(z).
\label{614}
\eeq
It follows that, in units such that $M_5^3=1$,
\beq
A_{\nu_0}=(4\pi M_p^2)^{-1}=2G,
\label{615}
\eeq
where $G$ is the Newton coupling constant. The scalar and dilaton
fluctuations (\ref{324}), using the static solutions (\ref{610}), can then
be  written in the form:
\bea
&&
\varphi =-{GM\over r}\left[1+ {2 \a_2\over 1+12 \a_2^2} \left({Q\over
M} + 2 \a_2\right) + {4\over 3}B_{\nu_0} \left(1\over
kr\right)^{2\nu_0-2}\right], \nonumber\\
&&
\psi =-{GM\over r}\left[1-{2 \a_2\over 1+12 \a_2^2} \left({Q\over
M} + 2 \a_2\right) + {2\over 3}B_{\nu_0} \left(1\over
kr\right)^{2\nu_0-2}\right], \nonumber\\
&&
\Ga =-{GM\over r}\left[{4 \a_2\over 1+12 \a_2^2} \left({Q\over
M} + 2 \a_2\right) + {2\over 3}B_{\nu_0} \left(1\over
kr\right)^{2\nu_0-2}\right], \nonumber\\
&&
\chi =-{GQ\over r}\left[{2 \over 1+12 \a_2^2} \left(1+ 2\a_2
{M\over Q}\right) + {2}B_{\nu_0} \left(1\over
kr\right)^{2\nu_0-2}\right].
\label{616}
\eea
It should be noted that the short-range corrections induced by the
massive scalar modes have the same qualitative behaviour as in the
tensor case, discussed in \cite{27}, in spite of the fact that the massive
scalar modes have different spectra. The dimensional decoupling
(i.e., the suppression of the higher-dimensional corrections) is thus 
effective  for all scales of distance $r$ such that $kr \gg 1$, where
$k=(M_5^3/M_p^2)(\nu_0-1)^{-1}$ is the mass scale relating the five-
and four-dimensional gravitational constants,  in our class  of
backgrounds.

The limiting case $\Da=-8/3$, i.e.  $\a_2=0$ and $\nu_0=2$,
corresponds to a pure AdS$_5$ background, if there are no scalar
charges on the brane. In that case $\om_2$  
exactly corresponds
to the  dilaton fluctuation $\chi$ (see eq. (\ref{323})), and can be
consistently set to zero (toghether with the dilaton background) if we
want to match, in particular, the ``standard" brane-world configuration
originally  considered by Randall and Sundrum \cite{9a}. In this limit,
$B_{2}=1/2$, and we exactly recover previous results for the effective
gravitational interaction on the brane \cite{27a}, i.e.
\beq
\varphi= -{GM\over r} \left(1+ {2\over 3 k^2r^2}\right),
~~~~~~ \psi= -{GM\over r} \left(1+ {1\over 3 k^2r^2}\right).
\label{617}
\eeq

The massless-mode  truncation reproduces in this case
the static, weak field limit of linearized  general relativity 
(note, however, that for $Q \ne 0$ there is no way to get rid of the
long-range scalar interactions). The massive
tower of scalar fluctuations, however, induces deviations from Einstein
gravity already in the static limit (as noted in \cite{27a}), and is the
source of a short-range force due to the ``breathing" of the fifth 
dimension, 
\beq
\Ga= - {GM\over 3 r}{1\over (kr)^2},
\label{618}
\eeq
even in the absence of bulk scalar fields, and of scalar
charges  for the matter  on the brane.

In a more general gravi-dilaton background ($\Da \not= -8/3$,
$\a_2\not=0$), the static expansion (\ref{616}) describes an effective
scalar-tensor interaction on the brane, which is potentially dangerous
for the brane-world scenario, as it contains not only short-range
corrections, but also long-range scalar deviations from general
relativity (and, possibly, violations of the
Einstein equivalence principle), even in the interaction of ordinary
masses, i.e. for $Q=0$ (similar results have   been recently obtained
also in the context of a multibrane scenario \cite{39}). 
This seems to offer an
interesting window to investigate the effects of the bulk geometry on
the four-dimensional  physics of the brane.

\section{Conclusions}
\setcounter{equation}{0}
\def\theequation{\thesection.\arabic{equation}}

In this paper we have analysed the full set of coupled equations
governing the evolution of scalar fluctuations in a dilatonic brane-world
background, supporting a flat 3-brane rigidly located at the fixed point
of $Z_2$ symmetry. We have diagonalized the system of dynamical
equations, and found four decoupled but self-interacting variables
representing, in a five-dimensional bulk, the four independent degrees
of freedom of scalar excitations of the gravi-dilaton background.

We have then restricted our discussion to the class of background
solutions characterized by a  brane of positive tension, by a
decreasing warp-factor as we move away from the brane, and by the absence of
naked singularities \cite{27}. Such a class of backgrounds can be
characterized by a real parameter $\Da$ ranging from $-8/3$ to $-2$
or, alternatively, by a real parameter $\nu_0$ ranging from $2$ to
$+\infty$. The limiting case $\Da =-8/3$, $\nu_0=2$, corresponds to the
``pure-gravity" AdS$_5$ background \cite{9a}.

 We have presented the exact solutions of
the canonical perturbation equation for all the scalar degrees of
freedom, and we have discussed, in this class of backgrounds, the
effects of their massless and massive spectrum for the scalar
interactions on the brane, taking into account the appropriate parity
under $Z_2$ symmetry, the normalization condition for the bound
states of the spectrum, and the first-order differential constraints
arising from the dynamics of scalar perturbations.

We have found that, for all backgrounds, there is  one
propagating massless mode localized on the brane,  associated with a
long-range dilatonic interaction in four dimensions.
Only very exceptionally (i.e. for a RS background and vanishing  
dilatonic charges) this ``fifth force" disappears. This interaction is  
always
affected by  higher-dimensional corrections, due to the scalar massive
modes that are not confined on the brane and can freely propagate in
the bulk space-time. The amplitudes of  such massive modes are
constrained by the  scalar perturbation equations and, in general, only
two amplitudes can be independently assigned.
The scalar fluctuation spectra are in general different from the
corresponding spectra of the spin-2 degrees of freedom. In the
weak and static limit, however, the leading-order short-range
corrections to the scalar force have the same radial dependence as in
the case of pure tensor interactions.

In particular, we have found a non-trivial massive spectrum of scalar
metric fluctuations even in the pure Randall--Sundrum scenario with
AdS$_5$ metric \cite{9a}. In more general backgrounds we have found
that there are also scalar contributions to the long-range interactions
of two massive bodies, even in the absence of specific ``dilatonic"
charges, with a resulting effective scalar-tensor interaction on the
brane. In this sense, the
bulk geometry seems  to affect not only the radial dependence,
but also the spin  content of the effective gravitational forces.
We believe that this effect is potentially
interesting for further applications of (and constraints on)
 the brane-world scenario.

\section*{Acknowledgements}
It is a plesure to thank P. Bin\'etruy, M. Giovannini, D. Langlois and C.
Ungarelli for helpful conversations.
G.V. wishes to acknowledge the support of a ``Chaire Internationale
Blaise Pascal", administered by the ``Fondation de L'Ecole Normale
Sup\'erieure'', during most of this work.


\newpage
\begin{thebibliography}{99}
\newcommand{\bb}{\bibitem}

\bb{1}P. Horava and E. Witten, Nucl. Phys. B {\bf 460}, 506 (1996).

\bb{2}P. Horava and E. Witten, Nucl. Phys. B {\bf 475}, 96 (1996).

\bb{4}A. Lukas, B. A. Ovrut, K. S. Stelle and D. Waldram, Phys. Rev.
D{\bf 59}, 086001 (1999).

\bb{5}A. Lukas, B. A. Ovrut, K. S. Stelle and D. Waldram, Nucl. Phys.  
B{\bf 552}, 246 (1999).

\bb{9}I. Antoniadis,  Phys. Lett. B {\bf 246}, 377 (1990).

\bb{6}N. Arkani-Hamed, S. Dimopoulos and G. Dvali, Phys. Lett. B
{\bf 429}, 263 (1998).

\bb{7}I. Antoniadis, N. Arkani-Hamed, S. Dimopoulos and G. Dvali, Phys.
Lett. B {\bf 436}, 257 (1998).

\bibitem{3}V. Rubakov and M. E. Shaposhnikov, Phys. Lett.
{\bf 125B}, 139 (1983).

\bb{8} L. Randall and R. Sundrum,  Phys. Rev. Lett.
{\bf 83}, 3370 and 4690  (1999).

\bb{9a} L. Randall and R. Sundrum,  Phys. Rev. Lett.
{\bf 83}, 4690  (1999).

\bb{10}R. B. Abbott, B. Bednarz and S. D. Ellis, Phys. Rev. D
{\bf 33}, 2147 (1986).

\bb{11}J. Garriga and E. Verdaguer, Phys. Rev. D
{\bf 39}, 1072 (1989).

\bb{12}L. Amendola, M. Litterio and F. Occhionero, Phys. Lett. B
{\bf 237}, 348 (1990).

\bb{13}M. Gasperini and M. Giovannini, Phys. Rev. D
{\bf 47}, 1519 (1993).

\bb{14}M. Gasperini and M. Giovannini, Class. Quantum Grav.
{\bf 14}, 735 (1997).

\bb{15}M. Giovannini, Phys. Rev. D
{\bf 55}, 595 (1997).

\bb{16}C. van de Bruck, M. Dorca, R. H. Brandenberger and A. Lukas,
Phys. Rev. D  {\bf 62}, 123515 (2000).


\bb{17}C. Csaki, J. Erlich, T. J. Hollowood and T. Shirman, Nucl. Phys. B
{\bf 581}, 309 (2000).

\bb{TV} T. R. Taylor and G. Veneziano, Phys Lett. B {\bf 213}, 459 (1988).

\bb{tests} E. Fischbach and C. Talmadge, Nature {\bf 356},  
207 (1992).

\bb{DP} T. Damour and A. M. Polyakov, Nucl. Phys. B {\bf 423}, 532 (1994).

\bibitem{18} D. Langlois,  Phys. Rev. D {\bf 62}, 126012 (2000).

\bibitem{19} D. Langlois,  R. Maartens, M. Sasaki and D. Wands,
Phys. Rev. D {\bf 63}, 084009 (2001).

\bb{20}B. Grinstein, D. R. Nolte and W. Skiba, hep-th/0005001.

\bb{21}N. Deruelle, T. Dolezel and J. Katz, hep-th/0010215.

\bb{22}U. Gen and M. Sasaki, gr-qc/0011078.

\bb{23}A. Neronov and I. Sachs, hep-th/0011254.

\bb{24}C. van de Bruck and M. Dorca, hep-th/0012073.

\bb{25} R. Maartens, gr-qc/0101059.

\bb{26}T. Boehm, R. Durrer and C. van de Bruck,  hep-th/0102144.

\bibitem{27} M. Cvetic, H. Lu and C.N. Pope,
Phys. Rev. D {\bf 63}, 086004 (2001).

\bb{27a} J. Garriga and T. Tanaka, Phys. Rev. Lett.  {\bf 84}, 2778 (2000).

\bibitem{28} M. J. Duff,
in {\em Second Conference on Gauge Theories, Applied Supersymmetry
and Quantum Gravity} (Imperial Coll. Press, London, 1996).

\bb{29}M. J. Duff, Class. Quantum Grav. {\bf 5}, 189 (1988).

\bibitem{30} M. Cvetic, S. Griffies and H. H. Soleng,
Phys. Rev. D {\bf 48}, 2613 (1993).

\bb{31}V. F. Mukhanov, H. A. Feldman and R. Brandenberger, Phys. Rep.
{\bf 215}, 203 (1992).

\bb{32} I. Y. Aref'eva {\em et al}, Nucl. Phys. B {\bf 590}, 273 (2000).

\bb{33}M. Abramovicz and I. A. Stegun, {\em Handbook of Mathematical
Functions} (Dover, New York, 1972).

\bb{34}A. Brandhuber and K. Sfetsos, JHEP {\bf 9910}, 013 (1999).

\bb{39}I. I. Kogan, S. Mouslopoulos, A. Papazoglou and L. Pilo,
hep-th/0105255.

\end{thebibliography}
\end{document}